%% file: main.tex
\documentclass[reprint, superscriptaddress, prl, footinbib]{revtex4-2}

\usepackage{amsmath}
\usepackage{amsfonts}
\usepackage{amssymb}
\usepackage{graphicx}
\usepackage[usenames, dvipsnames]{color}
\usepackage{natbib}
\usepackage{hyperref}

\definecolor{valentin}{RGB}{8, 133, 161}
\newcommand{\update}[1]{\textcolor{black}{#1}}

\definecolor{enrico}{RGB}{247, 122, 94}
\newcommand{\enrico}[1]{\textcolor{black}{#1}}

\begin{document}

\title{Imaging heat transport in suspended diamond nanostructures with integrated spin defect thermometers}

\author{V. Goblot}
%\thanks{These authors contributed equally to this work.}
\email{Contact author: valentin.goblot@epfl.ch}
\affiliation{Institute of Physics, Swiss Federal Institute of Technology Lausanne (EPFL), CH-1015 Lausanne, Switzerland} 
\affiliation{Center of Quantum Science and Engineering, Swiss Federal Institute of Technology Lausanne (EPFL), CH-1015 Lausanne, Switzerland
}

\author{\!\!$^{,\dagger}$~K. Wu} % dirty workaround
%\author{K. Wu}
\thanks{These authors contributed equally to this work.}
\affiliation{Institute of Physics, Swiss Federal Institute of Technology Lausanne (EPFL), CH-1015 Lausanne, Switzerland}
\affiliation{PROUD SA, Lausanne, Switzerland}

\author{E. Di Lucente}
\affiliation{Theory and Simulation of Materials (THEOS), and National Centre for Computational Design and Discovery of Novel Materials (MARVEL), École Polytechnique Fédérale de Lausanne, 1015 Lausanne, Switzerland}

\author{Y. Zhu}
\affiliation{Institute of Physics, Swiss Federal Institute of Technology Lausanne (EPFL), CH-1015 Lausanne, Switzerland} 

\author{E. Losero}
\affiliation{Institute of Physics, Swiss Federal Institute of Technology Lausanne (EPFL), CH-1015 Lausanne, Switzerland} 
\affiliation{Istituto Nazionale di Ricerca Metrologica (INRiM), Strada delle Cacce 91, Torino, 10135, Italy}

\author{Q. Jobert}
\author{C. Jaramillo Concha}
\affiliation{Institute of Physics, Swiss Federal Institute of Technology Lausanne (EPFL), CH-1015 Lausanne, Switzerland}

\author{N. Quack}
\affiliation{Institute of Microelectronics, University of Stuttgart, 70569 Stuttgart, Germany} 

\author{N. Marzari}
\affiliation{Theory and Simulation of Materials (THEOS), and National Centre for Computational Design and Discovery of Novel Materials (MARVEL), École Polytechnique Fédérale de Lausanne, 1015 Lausanne, Switzerland}
\affiliation{Laboratory for Materials Simulations, Paul Scherrer Institut, 5232 Villigen PSI, Switzerland}

\author{M. Simoncelli}
\email{Contact author: michele.simoncelli@columbia.edu}
\affiliation{Department of Applied Physics and Applied Mathematics, Columbia University, New York (USA)}
\affiliation{Theory of Condensed Matter Group of the Cavendish Laboratory, University of Cambridge, United Kingdom}

\author{C. Galland}
\affiliation{Institute of Physics, Swiss Federal Institute of Technology Lausanne (EPFL), CH-1015 Lausanne, Switzerland} 
\affiliation{Center of Quantum Science and Engineering, Swiss Federal Institute of Technology Lausanne (EPFL), CH-1015 Lausanne, Switzerland
}%a
\date{\today}

\begin{abstract}

Among all materials, mono-crystalline diamond has one of the highest measured thermal conductivities, with values above 2000 W/m/K at room temperature. This stems from momentum-conserving `normal' phonon-phonon scattering processes dominating over momentum-dissipating `Umklapp' processes, a feature that also suggests diamond as an ideal platform to experimentally investigate phonon heat transport phenomena that violate Fourier's law.
Here, we introduce dilute nitrogen-vacancy color centers as in-situ, highly precise spin defect thermometers to image temperature inhomogeneities in single-crystal diamond microstructures heated from ambient conditions.
We analyze cantilevers with cross-sections in the range from about 0.2 to 2.6 \textmu m$^2$, observing a \update{strong reduction of the cantilevers' conductivity as width decreases. We use first-principles simulations based on the linearized phonon Boltzmann transport equation and viscous heat equations to quantitatively predict the cantilevers' thermal transport properties, rationalizing how the interplay between intrinsic and extrinsic phonon scattering mechanisms determines the observed non-diffusive behavior.}
Our temperature-imaging method paves the way for the exploration of unconventional, non-diffusive heat transport phenomena in devices and nanostructures of arbitrary geometries. 
	
\end{abstract}

%\pacs{}
%\keywords{}

\maketitle

%Introduction

Carbon-based materials such as graphene, graphite or diamond present exceptional heat-conduction properties. 
Due to large phonon group velocities, they exhibit high thermal conductivity -- with diamond reaching record values at room temperature and below~\cite{Berman1975, Olson1993, inyushkin2018}. Moreover, the prevalence of normal (momentum-conserving) phonon-phonon scattering processes over Umklapp (momentum non-conserving) processes in these materials leads to phonon heat transport phenomena that violate Fourier’s law for macroscopic diffusive heat propagation~\cite{Chen2021}.
In graphene and graphite, ballistic transport in nanoscale devices~\cite{Bae2013, Nika2017} or hydrodynamic heat transport~\cite{cepellotti2015phonon, lee_hydrodynamic_2015, Huberman2019, Ding2022, Machida2020, Jeong2021, Huang2023} have been evidenced at or near room temperature. %, and recent theoretical work predicted that diamond should also exhibit hydrodynamic heat transport %deviating from Fourier's diffusion at room temperature~\cite{Simoncelli2020}.
\update{However, theoretical description of non-diffusive transport phenomena when both  ballistic (extrinsic) and  hydrodynamic (intrinsic) effects compete remains challenging, as it relies on phenomenological models that do not have a rigorous first-principles grounding~\cite{aharon2022direct, palm2024observation, Huang2023}.
In diamond,} the influence of phonon scattering at grain boundaries has been explored in polycrystalline samples~\cite{Sood2018} and unexpected size effects were recently reported in the thermal conductivity of microparticles~\cite{wang2024anomalously}, but experimental data on low-dimensional single-crystal diamond is lacking.
Understanding heat transport in nanostructured diamond will be valuable in view of its applications in optomechanics~\cite{Tao2014, Mitchell2016}, nonlinear optics~\cite{Hausmann2014}, quantum nanophotonics~\cite{moody2022}, as well as in heat management devices~\cite{Liu2015}.

A major challenge in the study of thermal transport at the nanoscale is to find appropriate temperature sensors. Traditional measurements of the heat conductivity rely on resistive sensors combined with advanced sample engineering~\cite{Bae2013, Machida2020} and cannot provide spatially-resolved temperature maps. \update{Scanning thermal microscopy offers excellent spatial resolution, but is invasive and requires challenging calibration of the probe-sample contact thermal resistance~\cite{gomes2015scanning, zhang2020review}.} Non contact optical thermometry techniques such as transient thermal gratings~\cite{Minnich2011, Huberman2019}, and time- or frequency-domain thermoreflectance~\cite{Regner2013, Hu2015, Tian2018} have been widely employed on simple planar geometries but require the deposition of a transducer material, typically a metal~\cite{wilson2012thermoreflectance}, and don't provide a direct temperature readout~\cite{huang2024graphite}. Other optical techniques~\cite{Brites2012, Quintanilla2018}, including Raman sideband thermometry~\cite{Balandin2008, Ghosh2008, Braun2022,elhajhasan2023optical}, offer good spatial resolution but rather poor temperature accuracy, typically not better than a few Kelvin.

In this work, we use the spin resonance of dilute color centers in the diamond lattice as precise in-situ temperature sensors to investigate heat transport properties of suspended diamond microstructures. \update{We then obtain quantitative agreement with theoretical predictions from multiscale first-principle modeling.}
Diamond hosts optically active spin defects, among which the negatively charged nitrogen vacancy (NV) center~\cite{Doherty2013}, which has been used as in-situ nanoscale thermometer~\cite{Neumann2013, Toyli2013, Kucsko2013}\update{, also in combination with a scanning tip~\cite{Laraoui2015}}.
Negatively charged NV centers are spin-1 atomic defects formed by a substitutional nitrogen atom neighboring a vacancy in the diamond lattice. Their spin can be initialized and readout by optical pumping and monitoring the photoluminescence intensity. The resonance frequency associated with the transition between $m_s=0$ and $m_s=\pm1$ spin states is temperature-dependent~\cite{Acosta2010}, as can be tracked with either continuous-wave or pulsed optically detected magnetic resonance (ODMR)~\cite{Barry2020}. NV centers are thus excellent thermometers, offering nanoscale spatial resolution and sub-$\mathrm{mK / \sqrt{Hz}}$ sensitivity~\cite{Wojciechowski2018} for steady-state temperature measurement near room temperature. 
Solid-state spin defects acting as temperature sensors are also found in a broad range of materials, including hBN~\cite{Gottscholl2021}, Si~\cite{Beaufils2018} and SiC~\cite{Kraus2014, Anisimov2016, Zhou2017}, making this method widely applicable.

\begin{figure}[t]
    \centering
    \includegraphics[width=0.75\linewidth]{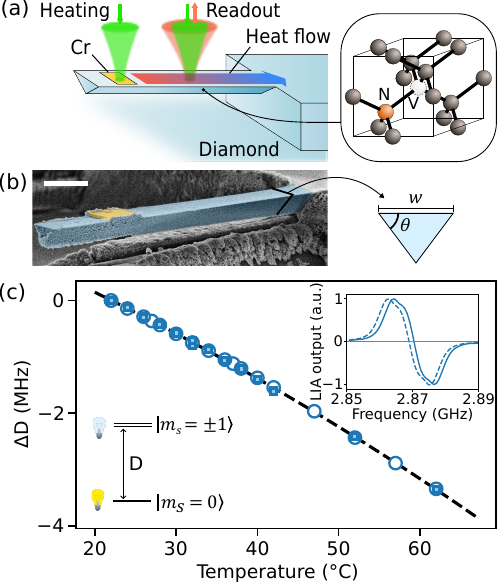}
    \caption{(a) Schematic representation of the experiment. Right inset: atomic structure of an NV center in the diamond lattice. (b) Scanning electron microscope (SEM) image with false colors for the diamond cantilever (blue) and Cr patch (yellow). Scale bar is 2~\textmu m. Right inset: triangular cantilever cross-section, with width $w$ and angle $\theta$. (c) ZFS shift $\Delta D (T) = D(T) - D(22 \mathrm{^\circ C})$  versus bath temperature, measured in bulk diamond (circles) and on a cantilever (squares). Lower inset: Energy levels of the NV center ground state. The ``bright'' $m_s=0$ state shows higher photoluminescence than the degenerate, ``dark'' $m_s=\pm1$ states, allowing to measure $D$ through ODMR. Top inset: lock-in ODMR spectra measured at $22~\mathrm{^\circ C}$ (solid line) and $62~\mathrm{^\circ C}$ (dashed).}%, showing a clear shift of the resonance.}
    \label{fig:Fig1}
\end{figure}

The principle of our experiment is summarized in Fig.\ref{fig:Fig1}a. 
We fabricate suspended micro-cantilevers out of single crystal diamond grown by chemical vapour deposition and enriched in $\mathrm{^{12}C}$ ($>99.95$\%), containing NV centers up to a concentration of 3~ppm (HiQuTe Diamond). Cantilevers 
%with lateral width $w$ ranging from 700~nm to 3~\textmu m 
are patterned using electron-beam lithography and subsequent etching, in two steps: first a step of vertical, oxygen-based reactive ion etching, followed by focused ion beam (FIB) milling at an angle $\theta = 53\text{\textdegree}$ from the diamond surface. This results in triangular cantilever cross-sections, as visible in Fig.~\ref{fig:Fig1}b. The cantilever acts as a one-dimensional channel for heat transport.
A small chromium patch is added at the cantilever tip to serve as a heat source upon absorption of a 515~nm laser focused onto it (referred to as heating laser) and
the bulk diamond acts as a heat sink.
%In the future, a UV-C laser could be used to heat diamond directly, as shown in other wide-gap materials~\cite{elhajhasan2023optical}.
Complete details on the sample fabrication can be found in the Supplementary Material (SM)~\cite{SupMat}.

Photoluminescence from NV centers is collected in confocal geometry, using a second 515~nm laser beam to address a large ensemble of NV centers (about $10^5$) located in the diffraction-limited spot. This second laser is referred to as readout laser. The temperature-dependent zero-field splitting (ZFS) between $m_s=0$ and $m_s = \pm1$ states is measured through continuous wave (cw) ODMR. We use a lock-in amplifier (LIA) combined with frequency modulation of the microwave to achieve high signal to noise ratio~\cite{Zhang2021robust}.
\update{
We first calibrate the variation of ZFS, $\Delta D (T) = D(T) - D(22 \mathrm{^\circ C})$, against a reference bath temperature $T$ applied to the entire sample. Fig.~\ref{fig:Fig1}c presents $\Delta D(T)$ measured both in bulk diamond and on cantilevers. Even though the absolute value of $D(T)$ is found to depend on position~\cite{SupMat}, possibly due to the presence of local strain, $\Delta D(T)$ is position-independent. Indeed, since our cantilevers are free-standing, we do not expect that heating induces any additional strain (except, maybe, very near its base).}
%Note that, at room temperature, the ZFS value is around 2.87 GHz. 
Our data is fitted by the model from Ref.~\cite{Cambria2023}, with only two fitting parameters (dashed line in Fig.~\ref{fig:Fig1}c). 

% The absolute value of the ZFS is found to depend on position~\cite{SupMat}, possibly due to the presence of local strain. Since our cantilevers are free-standing, we do not expect that heating induces any additional strain (except maybe very near its base). 
% We measure the change of ZFS against a reference temperature and present in Fig.~\ref{fig:Fig1}c the results for bulk diamond and a cantilever. 

The calibration is used to convert values of $\Delta D(T)$ to local temperature under the readout laser spot.
Noise analysis gives a typical temperature sensitivity of 0.27~$\mathrm{K / \sqrt{Hz}}$ for the temperature readout on the cantilever~\cite{SupMat}. %Note that the temperature sensitivity depends on the number of NV centers probed by the readout spot, it can thus degrade when addressing smaller cantilevers, but also vary with the NV density in the starting diamond substrate. 
The choice of cw-ODMR is a trade-off between sensitivity requirement and simplicity of experimental implementation. Even though more sophisticated NV-based thermometry schemes allow to reach better sensitivity, relying e.g. on pulse sequences~\cite{Neumann2013, Toyli2013} or on multi-frequency drive~\cite{Wojciechowski2018}, the dominant sources of uncertainties in our case come from the estimates of absorbed laser power and cantilever cross-section, as detailed below.

%Temperature profile in cantilevers
\begin{figure}[t]
    \centering
    \includegraphics[width=0.9\linewidth]{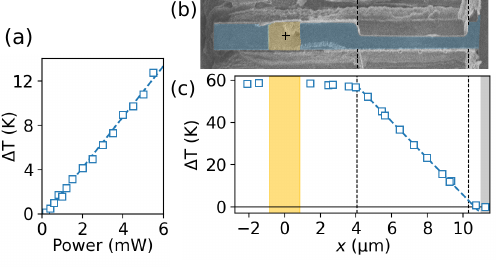}
    \caption{(a) Temperature shift $\Delta T$ at a fixed readout position on a  cantilever versus heating laser power $P_h$. Dashed line is a linear fit. (b) SEM image (top view) of a 1.53~\textmu m-wide cantilever (blue) with a narrow section (0.73~\textmu m) between the Cr patch (yellow) and the bulk.  Vertical dashed lines highlight the extremities of the narrow section. The black cross indicates the heating laser position (c) Temperature profile measured along the cantilever at $P_h = 3.0$~mW, by sweeping the position $x$ of the readout laser. The dashed blue line is a linear fit of $\Delta T(x)$ in the narrow section.}
    \label{fig:Fig2}
\end{figure}

\begin{figure*}[t]
    \centering
    \includegraphics[width=\linewidth]{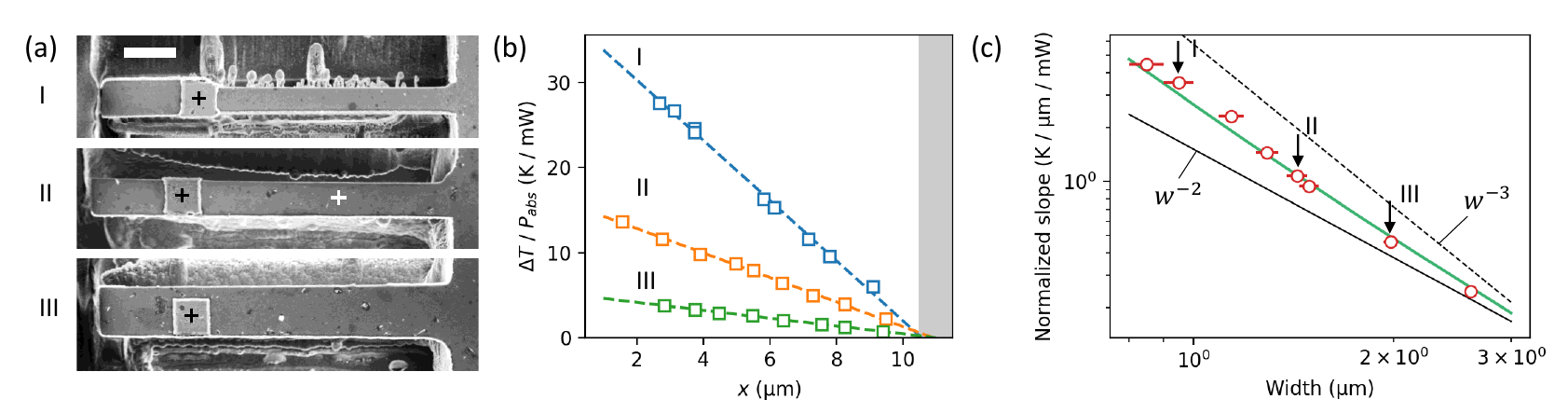}
    \caption{(a) SEM images, top view of cantilevers with width: $w = 0.95$~\textmu m (A), 1.43~\textmu m (B) and 1.98~\textmu m (C). The black cross indicates the heating laser position, corresponding to $x=0$ in (b). \update{The white cross in II indicates the readout position for the power scan in Fig.~\ref{fig:Fig2}a.} (b) Temperature profile measured in corresponding cantilevers, normalized by \update{the absorbed heating power $P_{abs} = \alpha P_h$.} (c) Normalized temperature gradient extracted from the profiles in (b) versus cantilever width $w$. \update{Black lines correspond to the $w^{-2}$ scaling predicted by Fourier's law (solid line) and the $w^{-3}$ scaling (dashed line) expected in the ballistic limit; green line is the first-principle prediction from the VHE.} }
    \label{fig:Fig3}
\end{figure*}

The heating laser is used to induce a heat flow in a cantilever.
Fig.~\ref{fig:Fig2}a presents the measured local temperature rise $\Delta T$ as a function of heating laser power $P_h$, for a readout spot located at a fixed position in the middle of the cantilever \update{(see Fig.~\ref{fig:Fig3}a,II)}. We observe a linear relation between $\Delta T$ and heating power.
%this behavior is in agreement with  Fourier's law predictions --- we stress that this behavior is necessary but not sufficient to ensure that Fourier's law is applicable, and we will analyze the size-dependence of the heat flux later.
% Move following paragraph to Sup Mat?
Standard deviations of residuals after a linear fit give a temperature error of $\pm0.25$~K~\cite{SupMat}. This value includes systematic errors due e.g to imperfect alignment of the heating laser on the chromium patch, explaining why it is slightly higher than the readout error ($\pm0.15$~K with our 3~s integration time).

Fixing the heating laser power, we measure the space-dependent temperature profile along the cantilever by sweeping the position of the readout laser. For each readout laser position, the ZFS values are measured without and with heating. To illustrate the method, the temperature profile for a cantilever of varying width is shown in Fig.~\ref{fig:Fig2}c. \update{Each section of constant width shows a linear temperature profile. The narrow section (from 4 to 10~\textmu m) has a much steeper temperature gradient than the wide section (0 to 4~\textmu m), while the temperature is constant on the free-hanging side ($-2$ to 0~\textmu m).}
In our experimental setting, the spatial resolution is diffraction-limited, the readout laser spot size being $\sim$0.44~\textmu m in diameter.

%We emphasize that the ZFS can also shift due to local strain or stress. This is for example visible in the evolution of ZFS along the cantilever (in the absence of heating), where a strong variation is observed near the cantilever base~\cite{SupMat}. However, since our cantilevers are free-standing, we do not expect that heating induces any additional strain except near its base.

% Sentence below moved to the SupMat
%We have also verified that the heating laser does not induce any change in ZFS when focused on other portions of the cantilever.

%Effect of size

 \update{To better understand this temperature profile,} we consider the expected behavior for standard diffusive heat transport. According to Fourier's law, in 1D, $q = - \kappa \, \partial T / \partial x$, with $q$ the local heat flux density and $\kappa$ the material's thermal conductivity, which depends slowly on temperature. 
 We confirmed that the Cr patch's temperature remains close to that of the underlying diamond~\cite{SupMat}, so that convective and radiative thermal losses can be neglected. In this case, $q = \alpha P_h / S$, with $P_h$ the heating laser power incident on the Cr patch, $\alpha$ the Cr patch absorption coefficient and $S = \tan (\theta) w^2/4$ the cantilever cross-section. For small $\Delta T$, Fourier's law predicts a linear temperature profile along the cantilever, compatible with the results from Fig.~\ref{fig:Fig2}.
However, having a linear temperature profile is a necessary but not sufficient condition for the validity of Fourier's law, since it also assumes thermal conductivity to be an intrinsic, size-independent material property, and consequently predicts a heat flux that scales linearly with the cantilever cross-section. 
Instead, in diamond, $\kappa$ has been predicted to decrease to approximately half of its bulk value at widths $\sim$0.7$\mu$m~\cite{li2012thermal, fugallo_ab_2013} -- a much stronger reduction than other materials for this channel size\cite{swinkels2015diameter,martin2009impact,hochbaum2008enhanced,li2003thermal,maire2017ballistic}. %However, the width dependence of thermal conductivity in diamond has not been experimentally demonstrated yet. 
We now show that a reduction of $\kappa$ is observed when reducing the lateral size of the cantilever.
We investigate cantilevers with different lateral width $w$, in the range 0.8-2.5~\textmu m, and identical length of 10~\textmu m.
The lower bound of this range originates from limitations in temperature sensitivity 
for cantilever widths below the diffraction limited spot size (due to the reduced number of probed NV centers). Conversely, the upper bound is due to the difficulty in fabricating suspended cantilevers with dimensions bigger than 3~\textmu m (due to the long etching time required).
We measure the temperature profile along each individual cantilever, adjusting $P_h$ such that the maximum temperature rise is comparable in all cases and remains below 40~K, within the calibration range of Fig.~\ref{fig:Fig1}. The results are presented in Fig.~\ref{fig:Fig3}, with panel (c) presenting the measured temperature gradient $\partial T / \partial x$, normalized by $\alpha P_h$, versus cantilever width $w$. In the case of purely diffusive thermal transport, \update{i.e., size-independent $\kappa$,} Fourier's law predicts that this normalized temperature gradient is inversely proportional to the cantilevers cross-section. Specifically, since all our cantilevers have triangular cross-section with same angle $\theta$,  Fourier's law predicts $\partial T / \partial x \propto w^{-2}$.  In contrast, our experiments \update{show a clear deviation  from this prediction for $w < 2$~\textmu m.
On the other hand, in the limit purely dominated by boundary scattering, thermal conductivity is expected to scale as $\kappa \propto w$\cite{ziman2001electrons}, corresponding to $\partial T / \partial x \propto w^{-3}$.
Our experiments are between these two regimes, indicating the necessity of accounting for both extrinsic (boundary) and intrinsic (normal and Umklapp) phonon scattering mechanisms.} 
%, with a temperature gradient displaying a scaling approximately proportional to $w^{-3}$ 

%This indicates that the probed range of cantilever width is within the non-Fourier regime, where non-diffusive effects are relevant and thermal conductivity depends on the sample's geometry (size).
%This is a strong indication that the probed range of cantilever width is already within the regime of finite-size limited thermal conductivity, and corresponds to a sharp drop in effective conductivity with channel size.

\begin{figure}[ht]
\centering
\includegraphics[width=\linewidth]{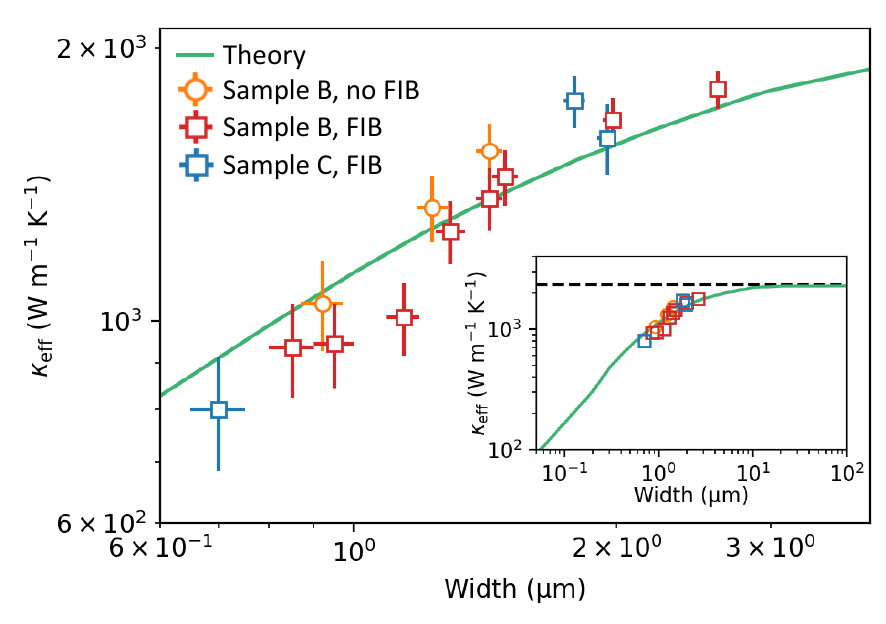}
\caption{\enrico{Effective thermal conductivity of diamond cantilevers as a function of width. Solid line is the prediction based on the VHE with parameters computed from first principles. 
Scatter points are experimental data, measured in two distinct diamond samples and fabricated either with or without FIB milling~\cite{SupMat}. The dashed line in the inset is the bulk conductivity.}
}
%results that account for Ga impurities introduced during fabrication and with concentration proportional to the surface-to-volume ratio (see text) are solid, while predictions that consider impurity scattering only from carbon isotopes (0.05\% of $^{13}$C) are dashed.
%Inset, cantilever cross-section with Ga impurities in proximity of the milled surfaces (penetration depth 0.01~\textmu m). The top surface is protected with a Ti layer during fabrication, hence not contaminated.
%$PR$ is defined as the ratio of Ga-contaminated area to total cross-section area.}
%causing $\kappa$ to decrease appreciably below its expected bulk value ($\rm{\sim2000W\,m^{-1}K^{-1}}$~\cite{slack1964thermal})
\label{fig:Fig4}
\end{figure}

We model thermal transport in our cantilevers using the viscous heat equations (VHE), 
two coupled mesoscopic heat-transport equations for temperature and phonon drift velocity~\cite{Simoncelli2020, dragasevic_viscous_2024} that generalize Fourier’s laws encompassing not only heat diffusion but also fluid-like behavior~\cite{Huberman2019,Ding2022,Machida2020,Jeong2021,PhysRevLett.125.265901}.
The VHE can be parametrized from first principles and allow to describe how micro-structured geometries influence thermal transport with accuracy comparable to microscopic models based on the full space-dependent solution of the linearized Boltzmann transport equation (LBTE)~\cite{raya-moreno_bte-barna_2022,raya-moreno_hydrodynamic_2022}, at much reduced computational cost~\cite{dragasevic_viscous_2024}. 
Therefore, we employ the VHE to investigate how the geometry of our cantilevers affects thermal transport, and we assess the relevance of viscous \update{and finite-size} effects by comparing the VHE predictions with those obtained from Fourier's law. 

We simulate the three-dimensional experimental geometry, imposing a temperature gradient $\Delta T = 10$~K around room temperature on $L=10$~\textmu m long cantilevers.
Parameters entering in the VHE or Fourier's law are computed from first-principles, accounting for third-order anharmonic phonon-phonon scattering~\cite{li2014shengbte,fugallo_ab_2013}, phonon-impurity scattering~\cite{ratsifaritana1987scattering}, as well as for the effects of rough cantilever's boundaries~\cite{li2012thermal} (see SM~\cite{SupMat} for details). 
\enrico{Importantly, the description of finite-size effects is based on the characteristic length scale of the problem, following
Casimir’s treatment of boundary scattering in small devices at low temperatures~\cite{carruthers1961theory}\footnote{We note that this is formally analogous to Bosanquet-type regularizations for the transport coefficients of rarefied fluids \cite{michalis_rarefaction_2010,dragasevic_viscous_2024}.}. In this regime, where boundary scattering occurs at a rate comparable to or higher than intrinsic anharmonic phonon scattering, the appropriate length scale to use is one-half of the smallest system dimension~\cite{cepellotti2017boltzmann}.}
Then, we solve the VHE or Fourier's law and determine the heat flux $\mathbf{q}$ from their solution. 
Finally, we evaluate the effective thermal conductivity as $\kappa_{\mathrm{eff}}{=} \frac{L}{\Delta T}[S^{-1}\iint_{S} \mathbf{q}\cdot d\mathbf{S}]$, where $S$ is the cantilever cross section, and $d\mathbf{S}$ the oriented surface element.

In Fig.~\ref{fig:Fig4} we compare the experimental values for $\kappa_{\rm eff}$, extracted from the measurements in Fig.~\ref{fig:Fig3} (see SM~\cite{SupMat} for details), against predictions obtained solving the VHE. 
\update{
We obtain good quantitative agreement between simulations and experiments. The striking deviation from diffusion, already discussed in Fig. 3, is visible as a sharp reduction of $\kappa_{\rm eff}$ away from the bulk value in Fig.~\ref{fig:Fig4}. A more detailed analysis of the strength of viscous hydrodynamic and ballistic effects reveals that the latter are the main contribution to the reduction of $\kappa_{\rm eff}$ (see SM~\cite{SupMat}). 
Heat transport in our cantilever is thus within the “apparent diffusion” regime~\cite{maassen2015steady}, in the sense that it is reasonably well described by a diffusive (Fourier-like) equation with a non-intrinsic thermal conductivity (i.e., depending on extrinsic factors such as the sample size). 
}

%We underline that the multiscale first-principles modeling developed in this work improves upon phenomenological models such as those used in Refs.\cite{aharon2022direct,palm2024observation}, since here the experiments are described through partial differential equations parametrized entirely from first principles. 

\update{
Fig. 4 also includes experimental data for cantilevers fabricated on distinct diamond samples, isotopically enriched in $^{12}$C and with the same NV concentration, but without relying on FIB (see SM~\cite{SupMat}). All datasets follow the same trend, confirming in particular that possible damage induced by FIB at the cantilevers' surface plays a negligible role in their transport properties.
}

\update{In conclusion, we have demonstrated that nearly one-dimensional diamond cantilevers with sub-micrometer widths and embedded NV centers exhibit clear signatures of non-diffusive thermal transport, manifested as an effective thermal conductivity with sub-ballistic size dependence. We have quantitatively explained these observations from a first-principles multiscale approach combining the linear-response solution of the Boltzmann transport equation and the mesoscopic viscous heat equations: sub-ballistic transport arises from the interplay between intrinsic phonon interactions---including both crystal-momentum-conserving``hydrodynamic'' scattering events as well as crystal-momentum-dissipating  Umklapp scattering events---and extrinsic effects determined by size and geometry.
This experiment sets the stage for investigating non-diffusive heat transport phenomena where intrinsic and extrinsic mechanisms coexist. First potential examples are nanostructured geometries that promote the formation of heat vortices~\cite{dragasevic_viscous_2024}, or induce thermal rectification~\cite{huang2024graphite}.}
\update{A second example is} investigating long-standing questions on the influence of sidewall roughness on phonon-boundary scattering~\cite{anthony_thermal_1991,Tian2024Phonon}, since by varying the etching method and chemical termination of the diamond surface one can engineer these extrinsic parameters. 
Third, this experimental setup can be pushed toward even smaller channels to address open questions in quantized phononic heat conductance~\cite{Polanco2023}. %, either using ultra-narrow pure diamond sections, or building hybrid devices, e.g. with carbon nanotubes bridging to diamond thermometer pads. 
Overall, our results rationalize non-diffusive, sub-ballistic heat transport in diamond, and more generally demonstrate the potential of solid state spin defects in ultrahigh thermal conductivity materials to explore heat transport phenomena in non-diffusive regimes.
%experimental platform we developed, consisting of NV centers in diamond nanostructures, enables the unambiguous detection of non-diffusive heat transport and offers multiple routes for new investigations.

%%%%%%%%%%%%%%%%%%%%%%%%%%%%%%%%%%%%%%%%%%%%%
\vspace{10mm}
\begin{acknowledgments}
\textit{Acknowledgments.} 
This project has received funding from the Swiss National Science Foundation through grant 198898 and 216406, EPFL Center for Quantum Science and Engineering through a Collaborative Research Fellowship, the European Union’s Horizon 2020 research and innovation programmes under the Marie Skłodowska-Curie grant agreement 945363 and under grant agreement 101112347, Project NerveRepack.
M. S. acknowledges support from: 
(i) the Kelvin2 HPC platform at the NI-HPC Centre (funded by EPSRC and jointly managed by Queen’s University Belfast and Ulster University); (ii) the UK National Supercomputing Service ARCHER2, for which access was obtained via the UKCP consortium and funded by EPSRC [EP/X035891/1]; (iii) Gonville and Caius College.  N.M and E.D. acknowledge support from the Swiss National Science Foundation (SNSF), through Grant No. CRSII5\_189924 (“Hydronics” project). The diamond cantilevers were fabricated
at the EPFL Center of MicroNanoTechnology (CMi) and we gratefully acknowledge the support of EPFL CMi.
\end{acknowledgments}

\bibliography{reference}

%%%%%%%%%%%%%%%%%%%%%%%%%%%%%%%%%%%%%%%%%%%%%

\newpage
\include{SupMat/SupMat}

\end{document}

%% file: SupMat/SupMat.tex
\onecolumngrid

\setcounter{equation}{0}
\setcounter{figure}{0}
\renewcommand{\theequation}{S\arabic{equation}}
\renewcommand{\thefigure}{S\arabic{figure}}

\section{Supplementary material}

\subsection{Sample preparation}

We use commercial (100) CVD single crystalline diamond substrate (Hiqute Diamond, Rouge-T12-100), enriched to above 99.95\% in $^{12}$C isotopes and containing an NV-rich layer (1--2 ppm) of $\sim3$~\textmu m thickness at the surface. We first perform preliminary acid cleaning: piranha solution (H$_2$SO$_4$:H$_2$O$_2$ 3:1) and hydrofluoric acid remove organics and metallic residuals, respectively. 
The substrate is then mounted on a silicon chip with 2~\textmu m thick wet oxide via QuickStick wax. We apply a reported non-contact ion beam polishing method to improve the surface roughness \cite{miNoncontactPolishingSingle2019}. 

To define the chromium patches, a 150~nm-thick Cr layer is evaporated on the NV-rich diamond surface. The patches are then defined using electron beam lithography followed by wet etching. 150 nm of hydrogen silsesquioxane (HSQ; XR-1541-006) is spin-coated, written with electron beam lithography, and subsequently developed in tetramethylammonium hydroxide (TMAH 25$\%$) as a mask. TechniEtch Cr01 from MicroChemicals ((NH$_4$)$_2$Ce(NO$_3$)$_6$+HClO$_4$) removes the unmasked Cr layer. 
Non-contact ion polishing is used to clean the sample surface, and then the patches are released by HF 1\%.

We then define the cantilevers, aligned to the Cr patches, using a second round of electron beam lithography followed by several etching steps. A thin titanium layer (40~nm) is evaporated on the surface. A 2~\textmu m-thick layer of HSQ (FOx-16) is spin coated above and exposed with electron beam lithography followed by the aforementioned development process. The Ti layer serves as a conductive and adhesive layer between diamond and HSQ.
A Chlorine-based reactive ion etching process is used to pattern the Ti layer (STS Multiplex ICP, 800 W ICP power, 150 W bias power, 10 sccm Cl$_2$ and 10 sccm BCl$_3$, 3 mTorr). To transfer the pattern from the HSQ hard mask to the diamond, a vertical etch is applied with directional oxygen plasma (STS Multiplex ICP, 400 W ICP power, 200 W bias power, 50 sccm O$_2$, 15 mTorr). Faraday cage angled etch is then used under the same plasma condition as the previous step to undercut the structures~\cite{burekHighQualityfactorOptical2014}. The hard mask is stripped in the HF 1\% bath. 

Since the Faraday cage angled etch technique restricts the lateral dimensions of suspended structures, we proceed further with focused ion beam (FIB) milling. This method allows greater flexibility in structuring, enabling us to precisely polish narrow cantilevers and undercut wider ones~\cite{graziosiSingleCrystalDiamond2018,riedrich-mollerOneTwodimensionalPhotonic2012}. A 200~nm Ti layer is deposited to shield the sample from gallium (Ga) implantation and to prevent drift during milling caused by charge accumulation. The milling is conducted with a Zeiss CrossBeam 540 SEM/FIB dual beams system using 30 kV Ga$^+$ ions on both sides of each cantilever. We apply cross-section mode with 1.5 nA beam current to sputter away most of the material and lower the ion current to 700 pA to realize better sidewall smoothness. Finally, cantilevers with finely polished sidewalls are revealed after removing the Ti layer in HF bath, as shown in Fig.~\ref{fig:Fig1}b.

\subsection{Experimental setup}

All experiments are carried out in our home-built confocal microscope. The diamond sample is mounted on a temperature-controlled stage, kept at a constant temperature of 22~$^\circ$C \update{(except for calibration measurements in Fig.~\ref{fig:Fig1})}. The output beam of a 515~nm laser is split into two beams, one for heating the chromium patch at the apex of the cantilever and the other one for reading out the NV centers' fluorescence all along the cantilever. The power of each beam can be controlled independently. The two beams are focused on different positions of the cantilevers with a high numerical aperture microscope objective (NA=0.9). The fluorescence is collected into a multi-mode fiber and detected with a multi-pixel photon counter (Hamamatsu C14455-1550GA).
% Complete details of the experimental setup can be found in~\cite{Babashah2023}.

Microwaves are delivered to the sample with a wire loop placed in close vicinity to the diamond sample. The microwave frequency is sinusoidally modulated at 5 kHz by the microwave generator (Rohde\&Schwarz SMB 100A) to be used in combination with the lock-in amplifier.
The readout parameters are chosen to optimize temperature sensitivity: the readout laser power is 0.15~mW, the microwave power is 41~dBm and the microwave frequency modulation depth is 5~MHz (corresponding to approximately half of the ODMR linewidth).

\begin{figure}[ht]
	\centering
	\includegraphics[width=0.7\linewidth]{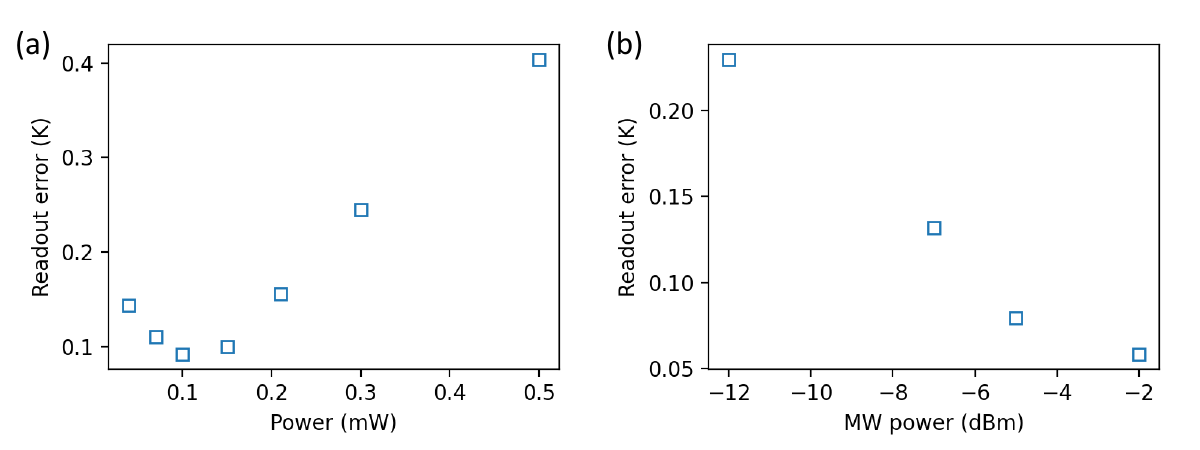}
	\caption{\update{(a) Influence of readout laser power on readout error, for a microwave power +41~dBm. The procedure for estimation of readout error is detailed in Fig.~\ref{fig:AllanDev}. (b) Influence microwave power on readout error, for a readout laser power of 0.15~mW. The reported microwave power values correspond to the output of the microwave generator, before +43~dBm amplification.}}
	\label{fig:LaserPower}

	\includegraphics[width=0.7\linewidth]{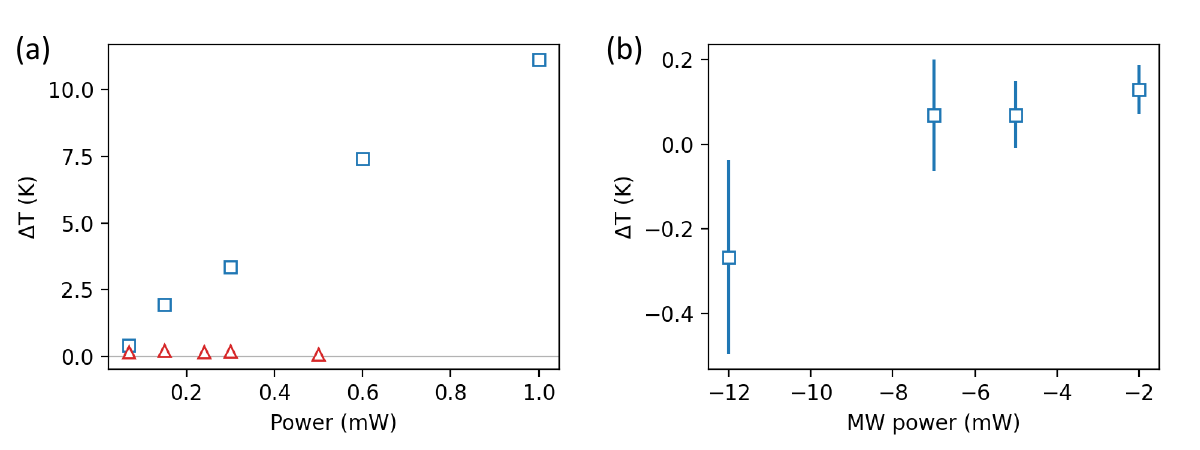}
	\caption{\update{(a) Influence of readout laser power on measured local temperature, on a cantilever fabricated with FIB milling (blue squares) and without FIB milling (red triangles), through Faraday cage angled etch only. (b) Influence of microwave power on measured temperature, on a cantilever fabricated with FIB milling. The reported microwave power values correspond to the output of the microwave generator, before +43~dBm amplification.}}
	\label{fig:Control}
\end{figure}

\update{
Fig.~\ref{fig:Control} shows how the above parameters affect the measured value of the zero-field splitting (ZFS). We find that, on cantilevers fabricated with FIB milling, the readout laser induces a small heating (blue squares). On the other hand, no laser-induced heating is measured on cantilevers fabricated only with Faraday cage angle etch, without the final FIB milling step (red triangles). We conclude that the layer contaminated with Ga ions during milling causes absorption of the readout laser light, which in turn generates local heating. 
Nevertheless, the small, $\sim 1$~K, heating caused by the readout laser with 0.15~mW power is not expected to affect the temperature profile measured in the presence of the heating laser on the metal patch: this contribution from the readout laser is removed when subtracting the ZFS values without the heating laser (see Fig.~\ref{fig:ZFSProfile}). The microwave power is found to have a negligible effect on the temperature readout.}
\update{Note that no static magnetic field is applied to the sample. This magnetic field-free scheme to measure the ZFS for thermometry is known to be insensitive to small magnetic field fluctuations~\cite{Zhang2021robust}.
}

\subsection{Independent study of the patch temperature}

\begin{figure}[ht]
    \centering
    \includegraphics[width=0.8\linewidth]{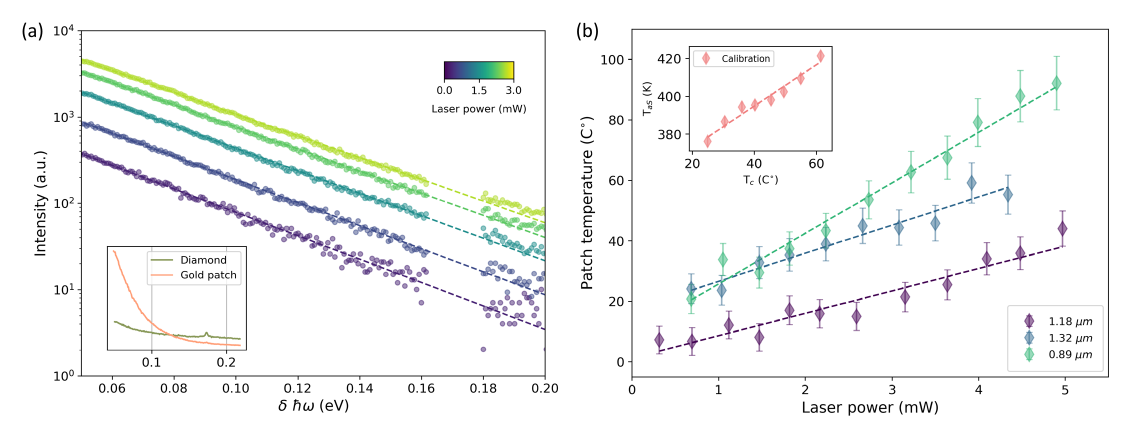}
    \caption{(a) Anti-Stokes emission spectra of Au patch at the tip of a $1.32$~\textmu m-wide cantilever, for different heating laser power. Dashed lines are fit using Eq.~\eqref{eq:aSfit}, allowing to obtain the temperature of hot carrier $T_{aS}$. (Inset) Comparison of emission spectra acquired on the diamond cantilever and Au patch, with identical experimental conditions. In the diamond spectrum, a Raman peak appears at $\delta\hbar\omega$=0.17 eV. 
    (b) Extracted patch temperature $T_p$ versus applied heating laser power, measured on three cantilevers with different widths. (Inset) Calibration of $T_{aS}$ as a function of sample base temperature $T_c$, performed on a large $20\,\mu m\times20\,\mu  m$ Au patch on bulk diamond to ensure the patch is at the same temperature as the temperature-controlled stage. Dashed line is a linear fit, used to convert $T_{aS}$ into patch temperature $T_p$ in the main panel.}
    \label{fig:aS_emssion}
\end{figure}

To confirm that the metal patch is well thermalized with the diamond cantilever, we measure the hot carrier anti-Stokes emission of the patch to get an independent estimate of its temperature. 
Luminescence in noble metals following photoexcitation is a complex process influenced by various mechanisms, including electron-electron scattering, electron-phonon coupling, multiphoton excitations, and surface plasmons \cite{caiAntiStokesEmissionHot2019,carattinoGoldNanoparticlesAbsolute2018}. Anti-Stokes emission corresponds to photons emitted at a higher energy than the energy of the laser incident on the material. 
%The anti-Stokes emission spectrum is proportional to the occupation number of the states that contribute energy. 
As detailed in previous studies, the anti-Stokes emission spectrum can be fit with Boltzmann statistics to obtain a first approximation of the effective temperature~\cite{caiAntiStokesEmissionHot2019,jollansEffectiveElectronTemperature2020}.
%The spectrum is given by:
%\begin{equation}
%I_{\rm{aS}}\propto P(\hbar \omega)D(E) {\rm{exp}}(\frac{-\hbar\delta \omega}{k_{\rm{B}}T_{\rm{aS}}})\label{eq:antiStokes}
%\end{equation}
%where $P(\hbar \omega)$ and $D(E)$ represent emission probability and a density of states respectively and $T_{\rm{aS}}$ represents the anti-Stokes temperature. The exponential term, or Boltzmann factor, merges the energy contribution from electrons and phonons. 
Under the assumption that both the photonic and electronic densities of states vary much slower than the Boltzmann factor close to the laser energy~\cite{cramptonJunctionPlasmonDriven2018,jollansEffectiveElectronTemperature2020}, we have the simple expression for the anti-Stokes intensity as a function of the Raman shift $\delta\omega$
\begin{equation}
I_{\rm{aS}}(\delta\omega)\propto A {\rm{exp}}(\frac{-\hbar\delta \omega}{k_{\rm{B}}T_{\rm{aS}}})
\label{eq:aSfit}
\end{equation}
from which we can determine the effective temperature $T_\mathrm{aS}$.

We measure the anti-Stokes emission spectrum on three different cantilevers, using 532~nm continuous wave excitation and a spectrometer. A notch filter rejects most of the reflected excitation laser. Collected spectra are shown in Fig.~\ref{fig:aS_emssion}a and can be well fitted with the exponential decay from Eq.~\eqref{eq:aSfit} over the $\delta \omega$ range that we can experimentally access. Note that we discard the energy window corresponding to the Raman peak of diamond, to avoid spurious residual contribution from the diamond cantilever. %that can be present due to the spot size comparable with the patch size.
We extract the anti-Stokes temperature $T_\mathrm{aS}$ from the fit. This experiment was performed on a patch made of gold (Au) instead of Cr, but with a few nm-thin Cr layer used as adhesion layer between Au and diamond. Hence, aside from the absorption coefficient of the patch $\alpha$, no major qualitative difference is expected in the metal-to-diamond thermalization efficiency.

The relationship between $T_\mathrm{aS}$ and the lattice temperature is in general not straightforward, specifically under continuous wave illumination~\cite{Dubi2019Hot}. 
Hence, we perform a calibration measurement to relate the extracted $T_\mathrm{aS}$ to the patch temperature, using the temperature-controlled stage on which the diamond sample is mounted (see inset of Fig.~\ref{fig:aS_emssion}b). We find that the relation between $T_{aS}$ and the controlled patch temperature $T_p$ can be well described using a linear fit: $T_{aS}=a T_p+b$. We can therefore determine the temperature of metal patches versus heating laser power, on cantilevers with different widths, as shown in Fig.~\ref{fig:aS_emssion}b. Although this method does not grant high precision (note also that the extracted zero-laser-power temperature is significantly lower on the 1.18~\textmu m cantilever), this measurement proves that the metal patch temperature increases linearly when heating laser is applied. Temperature estimates with anti-Stokes emission are also found to be of the same order as values presented in the main text on diamond cantilevers for similar heating power, meaning that the interface thermal resistance between the patch and diamond is sufficiently small. 

\update{
We can estimate convective and radiative thermal losses from the metal patch using the above temperature measurements.
For the range of applied laser power, we get patch temperature $T \leq 100~\mathrm{^\circ C}$. We use this value to get an upper-bound estimate of radiative and convective losses.
For radiative heat transfer, we use the Stefan-Boltzmann law for black-body radiation: 
\begin{equation}
	Q_{rad} = \epsilon \sigma T^4
\end{equation} 
where $\sigma$ is the Stefan-Boltzmann constant and $\epsilon$ is the emissivity of the body. Even for a perfectly black body, i.e., $\epsilon=1$, we get an upper bound $Q_{rad} = 1.5 .10^{-9}$~W.
Convective heat transfer can be estimated as 
\begin{equation}
	Q_{conv} = h A (T – T_{air})
\end{equation} 
with $A$ the patch area and $T_{air}$ the ambient air temperature. Considering a value $h = 50~\mathrm{W.m^{-2}.K^{-1}}$ of the heat transfer coefficient for free convection with air (which is on the higher end of values reported in the literature), $A = 1$~\textmu m$^2$ and $T_{air} = 20~\mathrm{^\circ C}$, we get an estimated $Q_{conv} = 4.10^{-9}$~W.
Both estimates for $Q_{rad}$ and $Q_{conv}$ are 5 to 6 orders of magnitude lower than the applied heating laser power.
Our main conclusion is thus that convective and radiative thermal losses from the metal patch to the surrounding can safely be neglected. The same is also true for radiative and convective losses in the diamond cantilever itself (it has even lower temperature and its outer surface is only 10 to 100x bigger than the patch). 
}

\subsection{Experimental estimate of the effective thermal conductivity}

The effective thermal conductivity of our cantilevers is defined as $\kappa_{{\rm{eff}}} = q \left( \mathrm{d}T / \mathrm{d}x\right)^{-1}$. The temperature gradient $\mathrm{d}T / \mathrm{d}x$ is directly extracted from the measured temperature profile. We only need to estimate the heat flux density $q$ in order to obtain $\kappa_{{\rm{eff}}}$. 
As introduced in the main text, we have $q = \alpha P_h / S$. The heating laser power incident on the Cr patch $P_h$ is measured directly. 
The Cr patch absorption coefficient $\alpha$ is estimated by measuring its reflection coefficient $r$ and assuming negligible transmission and scattering, i.e. $\alpha = 1 - r$. %Absence of transmission is justified by the 100~nm thickness of the Cr layer and negligible scattering is expected due to the large 0.9 NA of the microscope objective.
Finally, the cantilever cross-section is $S=\tan (\theta) w^2/4$, where both the width $w$ and cross-section angle $\theta$ are estimated from SEM images of the different cantilevers, see for example \update{Fig.~\ref{fig:Fig1}b and} Fig.~\ref{fig:Fig3}a of the main text. 
\update{Note that the explicit expression for $\kappa_{\rm{eff}}$ as a function of the above experimental parameters, specifically $w$ and $\mathrm{d}T / \mathrm{d}x$, is:
\begin{equation}
\kappa_{\rm{eff}} = \frac{4 \alpha P_h}{\tan (\theta) w^2 \left( \mathrm{d}T / \mathrm{d}x\right)}
\end{equation}}

We conservatively assume an uncertainty of 50~nm in the extracted width $w$ from SEM images. This uncertainty needs to be properly taken into account in the uncertainty of $\kappa_{{\rm{eff}}}$, together with the uncertainty on the temperature gradient. We find that the width uncertainty is the dominant factor, explaining the significantly bigger error bars in $\kappa_{{\rm{eff}}}$, see Fig.~\ref{fig:Fig4}, compared to the temperature gradient in Fig.~\ref{fig:Fig3}c.

\subsection{Additional figures}

\begin{figure}[ht]
    \centering
    \includegraphics[width=\linewidth]{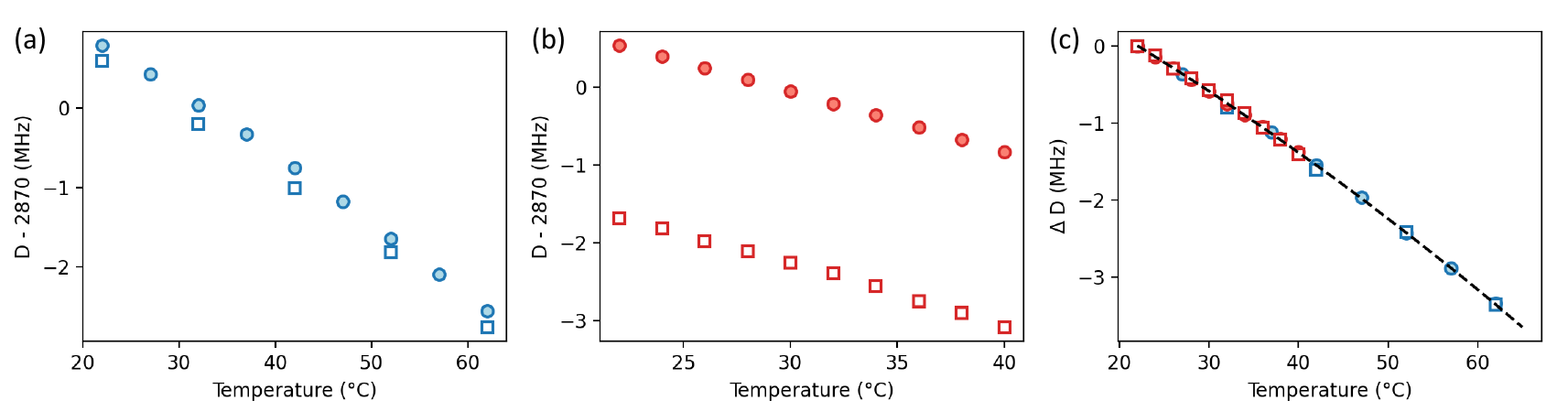}
    \caption{Detailed temperature calibration data from Fig.~\ref{fig:Fig1}c. (a, b) Raw ZFS value versus temperature measured in the bulk (circles) and on a cantilever (squares), for (a) sample A: a diamond with natural abundance (i.e 1.1\%) of $^{13}$C and low NV concentration of 10~ppb and (b) sample B: a diamond isotopically enriched to above 99.95\% $^{12}$C and NV concentration of 3~ppm. (c) Corresponding ZFS shift $\Delta D = D(T) - D(\mathrm{22~^\circ C}$ versus temperature from (a,b), with the same colors and symbols as in (a, b). Note that this is data is the same as in Fig.~\ref{fig:Fig1}c, but the colors clearly differentiate the two diamonds, highlighting that the temperature-induced ZFS shift is largely independent of the diamond hosting the NV centers. Note also that all other measurements presented both in the main text and Supplementary Materials are obtained with isotopically enriched diamond (samples B and C). Black dashed line in (c) is a fit with the expression from Ref.~\cite{Cambria2023}: $D(T) = D_0 + c_1 n_1 + c_2 n_2$, with $n_i = 1 / (e^{\Delta_i/k_bT} - 1)$ the phonon population in mode $i$, with energy $\Delta_i$. We use the same two mode energy as in~\cite{Cambria2023}, but keep the constants $c_{1, 2}$ as fitting parameters (here, $c_1 = -32.6$~MHz and $c_2 = -630$~MHz). Note that the fit gives a linearized temperature shift $dD/dT=-78~\mathrm{kHZ/K}$ around 30~$^\circ$C, corresponding to the widely reported value in the literature for room temperature.}
    \label{fig:TemperatureCalibration}
\end{figure}

\begin{figure}[ht]
    \centering
    \includegraphics[width=0.7\linewidth]{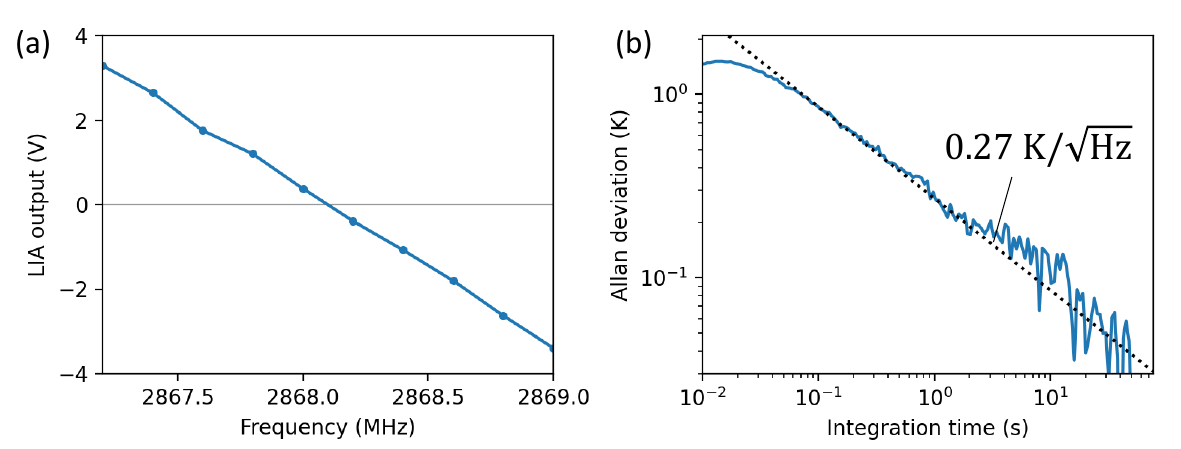}
    \caption{Estimation of temperature sensitivity. (a) Lock-in ODMR spectrum, measured on a  cantilever and at 22~$^\circ$C. Compared to the top inset in Fig.~\ref{fig:Fig1}c, here the microwave frequency range is limited to values around the NV resonance (i.e close to the ZFS value). A linear fit gives a slope of $-3.7~\mathrm{V/MHz}$, which characterizes the lock-in amplifier (LIA) output response to any ZFS shift. (b) Allan deviation of the LIA output, for the same experimental conditions as in (a) and a central microwave frequency of 2868.0 MHz. The conversion from LIA output noise to temperature noise is done by using the slope extracted from panel (a) and a linear approximation $dD/dT=-78~\mathrm{kHZ/K}$ (see discussion in the caption of Fig.~\ref{fig:TemperatureCalibration}). We get a temperature readout sensitivity of 0.27 K.Hz$^{-1/2}$. Note that the integration time in the 10-point spectrum shown in (a) is 3~s per point, yielding a temperature readout accuracy below 0.15~K.}
    \label{fig:AllanDev}
\end{figure}

\begin{figure}[ht]
    \centering
    \includegraphics[width=0.7\linewidth]{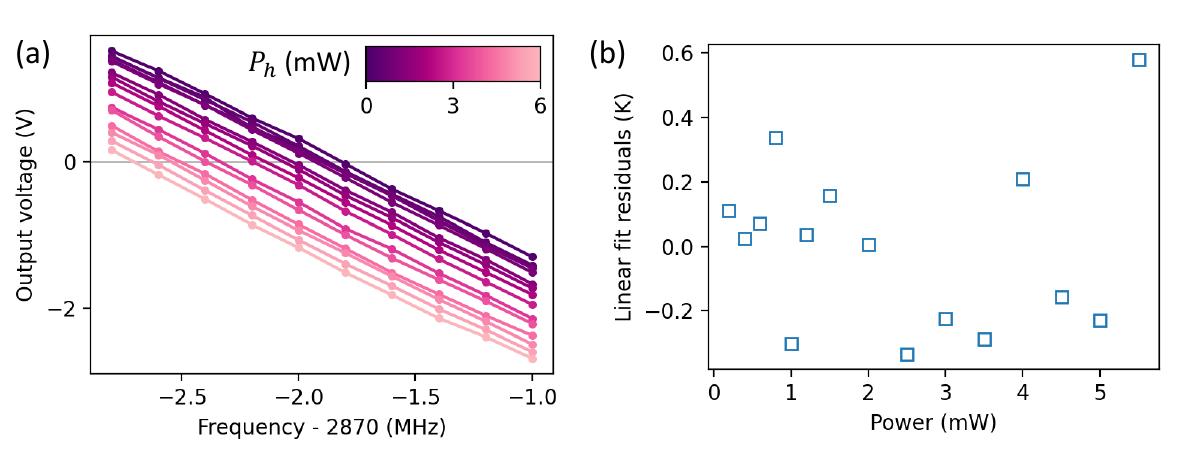}
    \caption{Detailed power scan data of Fig.~\ref{fig:Fig2}a. (a) Complete set of LIA ODMR spectra measured for increasing heating power $P_h$, as indicated by the color of each line. (b) Difference between the measured temperature shift and the linear fit of the power scan from Fig.~\ref{fig:Fig2}a. The standard deviation of these fit residuals is 0.25~K, corresponding to the temperature accuracy that includes systematic errors due e.g to imperfect alignment of the heating laser on the Cr patch. It is thus slightly higher than the readout error derived from the Allan deviation in Fig.~\ref{fig:AllanDev}.}
    \label{fig:PowerScan}
\end{figure}

\begin{figure}[ht]
    \centering
    \includegraphics[width=0.7\linewidth]{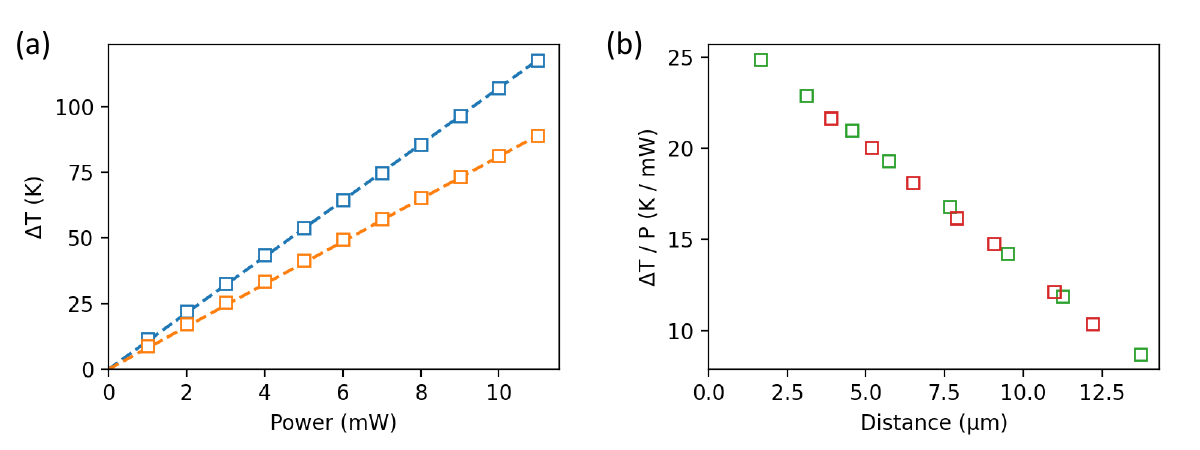}
    \caption{\update{(a) Power scans measured for a distance between metal patch and readout laser of $x = 2.47$~\textmu m (blue squares) and $x = 6.49$~\textmu m (orange), on a 1.23~\textmu m wide, 20~\textmu m long cantilever. (b) Temperature profile, normalized by absorbed heating power $\alpha P_h$, measured along the same cantilever, for heating laser power $P_h = 4.44$~mW (green squares) and $P_h = 10.0$~mW (red). These measurements confirm that temperature increases linearly versus heating power for all range of power probed, independent of the readout position on the cantilever.}}
    \label{fig:PowerComparison}
\end{figure}

\begin{figure}[ht]
    \centering
    \includegraphics[width=0.35\linewidth]{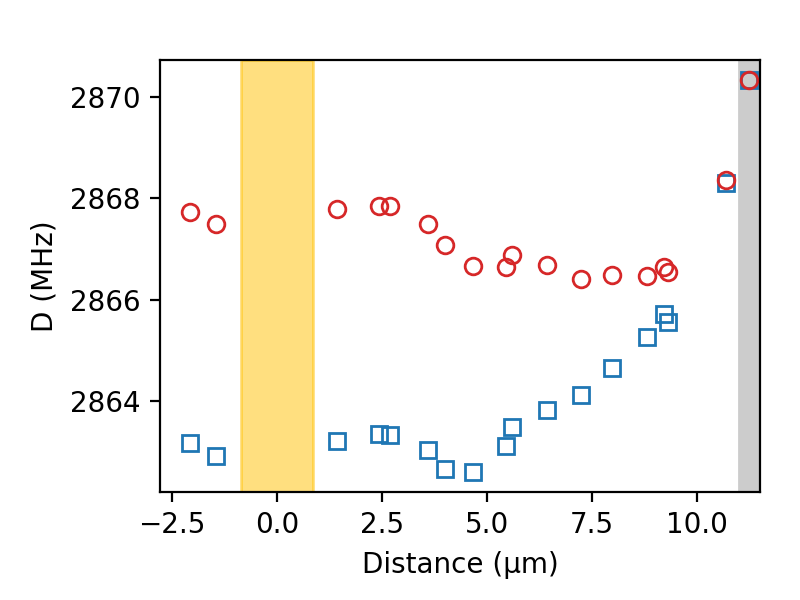}
    \caption{ZFS profile along the cantilever from Fig.~\ref{fig:Fig2}b, measured without heating, $P_h=0$, (red circles) and with heating, $P_h = 3.07$~mW (blue squares). Even in the absence of heating, strong variations of the ZFS along the cantilever are observed, attributed to local strain or stray electric fields caused by surface charges. The corresponding temperature profile shown in Fig.~\ref{fig:Fig2}c is obtained by subtracting the two measurements, and converting the resulting $\Delta D$ to temperature shift $\Delta T$ using the calibration from Fig.~\ref{fig:Fig1}(c).}
    \label{fig:ZFSProfile}
\end{figure}

\begin{figure}[ht]
    \centering
    \includegraphics[width=0.5\linewidth]{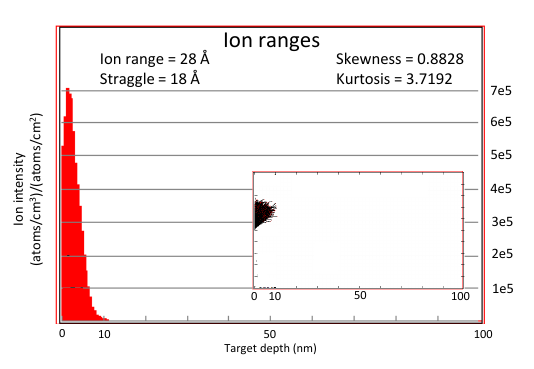}
    \caption{Monte Carlo simulations of 30~kV gallium ions coming under a polishing angle (1${^\circ}$) to the diamond\cite{Ziegler1985}. The SRIM simulation predicts that the depth of the implantation is of the order of 10~nm. The inset shows the collision plot.}
    \label{fig:Monte_Carlo}
\end{figure}

\clearpage
\subsection{Theoretical analysis}

An established approach to describe thermal transport in crystals from first principles is the linearized phonon Boltzmann transport equation (LBTE) \cite{peierls1929kinetischen,peierls1996quantum,ziman2001electrons}:
\begin{equation} \label{lbte}
	\frac{\partial n_{\nu}(\boldsymbol{r},t)}{\partial t}+\boldsymbol{v}_{\nu}\cdot\frac{\partial n_{\nu}(\boldsymbol{r},t)}{\partial\boldsymbol{r}}=\left.\frac{\partial n_{\nu}(\boldsymbol{r},t)}{\partial t}\right|_{{\rm{col.}}},
\end{equation}
where $n_{\nu}$ is phonon distribution ($\nu\equiv\boldsymbol{q}s$, with $\boldsymbol{q}$ being the phonon wavevector and $s$ the phonon mode index). The LBTE is a microscopic model, which describes the  propagation and scattering of phonons in the presence of a temperature gradient. The LBTE can be parametrized from first principles, and its linear-response solution allows to predict the thermal conductivity.
We note, in passing, that the LBTE has been recently generalized to account not only for the ability of phonons to propagate particle-like, but also for their capability to tunnel wave-like ~\cite{simoncelli2019unified,simoncelli2022wigner,di2023crossover}. In diamond, propagation dominates over tunneling and the LBTE accurately describes thermal transport~\cite{Simoncelli2020}. \update{Moreover, since in diamond momentum-conserving ``normal'' phonon scattering processes have higher (or comparable) strength than momentum-non-conserving  "Umklapp" processes, a solution of the LBTE accounting for the full scattering matrix is necessary to properly capture heat-transport properties~\cite{li2012thermal, fugallo_ab_2013}.}

The first-principles solution of the LBTE with full scattering matrix has the drawback of being computationally expensive, making it impractical to explore how nanostructure shape affects heat transport.
Therefore, to investigate thermal transport in cantilevers having different cross-section shapes and sizes, we rely on the mesoscopic viscous heat equations (VHE)~\cite{Simoncelli2020}. The VHE are two coupled partial differential equations for the mesoscopic temperature ($T$) and phonon drift-velocity  ($\boldsymbol{u}$) fields. They are obtained from coarse-graining the LBTE\cite{cepellotti2016thermal,Simoncelli2020}. The VTE encompasses both the diffusive (Fourier) and hydrodynamic regimes \cite{cepellotti2015phonon,cepellotti2016thermal} of thermal transport, as well as the crossover between them. 
It has been shown that the VHE rationalizes the experimental demonstration of hydrodynamic transport in graphite~\cite{Simoncelli2020}, and also yields predictions compatible with the space-dependent solution of the LBTE\cite{raya-moreno_bte-barna_2022,raya-moreno_hydrodynamic_2022} in complex-shaped devices~\cite{Dragasevic2023}.
In the steady-state regime, for a material with isotropic transport properties such as diamond, the VHE read:
\begin{equation} \label{VHE}
	\begin{cases}
		\alpha\nabla\cdot\boldsymbol{u}(\boldsymbol{R})=\kappa\nabla^{2}T(\boldsymbol{R})\\
		\beta\nabla T(\boldsymbol{R})-\mu\nabla^{2}\boldsymbol{u}(\boldsymbol{R})=-\gamma\boldsymbol{u}(\boldsymbol{R}),
	\end{cases}
\end{equation}
where $\mu$ is the phonon viscosity, the coupling coefficients $\alpha$ and $\beta$ arise from the energy-crystal momentum relation for phonons~\cite{Simoncelli2020}, while $\gamma$ accounts for the presence of momentum-dissipating scattering processes~\cite{Simoncelli2020}. We use capital $\boldsymbol{R}$ to emphasize the space variation at the mesoscopic scale. 

From Eq.~\eqref{VHE} one can derive energy and momentum balance equations in terms of a heat flux generalized to the hydrodynamic regime:
\begin{equation} \label{heat_flux}
	Q_{i}(\boldsymbol{R}) = -\kappa\sum_{j}\frac{\partial T(\boldsymbol{R})}{\partial R_{j}}+\alpha\sum_{j}u_{j}(\boldsymbol{R}),
\end{equation}
where $i$ labels Cartesian directions. 

The mesoscopic heat flux is then used to evaluate the effective thermal conductivity across the cantilevers. For a cantilever, taken along $x$ axis, of length $L$ and with an imposed temperature difference at its boundaries $\Delta T$, we get:
\begin{equation} \label{effective_kappa_em}
	\kappa_{{\rm{eff}}}=\bar{q}\frac{L}{\Delta T},
\end{equation}
where $\bar{q}=S^{-1}\iint_{S} \mathbf{q}\cdot d\mathbf{S}]$ is the average heat flux discussed in the main text.

All the parameters entering the VHE can be determined from the first-principles solution of the full LBTE (\textit{i.e.} accounting for the actual phonon band structure and full collision matrix). 
An exact solution of the LBTE has been discussed for the case of nanowires \cite{li2012thermal,li2014shengbte} under the assumption of diffusive boundaries. 
%The latter lead to breaking of translational symmetry perpendicular to the nanowire axis, translating into position-dependent phonon lifetimes in the LBTE. 
%In order to tackle the increased computational cost needed for solving such a space-dependent LBTE, an iterative solution based on nanowire cross-sectional averages has been developed \cite{li2014shengbte} by partially eliminating the two spatial dimensions perpendicular to the nanowire axis. 
This approach allows to account for the influence of boundary scattering on the conductivity of wires, providing an expression which depends only on the cross-section of the wire \footnote{In particular, Eq. 6 of Ref. \cite{li2012thermal} encompasses as special cases both thin films and thin wires, as discussed after Eq. 5 of Ref. \cite{chambers1950conductivity} and Eq. 11.2.9 of Ref. \cite{ziman2001electrons}}
and can be applied to the triangular cross-section nanowire studied in this work.
\update{Importantly, the description of finite-size effects is based on the characteristic length scale of the problem, following Casimir’s treatment of boundary scattering in small devices at low temperatures~\cite{carruthers1961theory}. In this regime, where intrinsic anharmonic phonon scattering occurs at a rate comparable to or higher than boundary scattering, the appropriate length scale to use is one-half of the smallest system dimension~\cite{cepellotti2017boltzmann}.}
After verifying with the established Casimir model \cite{fugallo_ab_2013} that boundary scattering has strong effects on the conductivity but has negligible effect on the other parameters entering the VHE (including the thermal viscosity $\mu$), we accounted for nanowire boundary scattering on the conductivity only and used bulk values (calculated in Ref.~\cite{Simoncelli2020}) for the other parameters entering in the VHE. This approximation is further justified a posteriori, since Fig.~\ref{fig:Fig_SI} shows that viscous effects are unimportant in the cases considered here.

%Finally, we discuss how we considered the Ga contamination in the vicinity of milled faces of the cantilevers (i.e. bottom faces, see inset of Fig.~\ref{fig:Fig4}). Since with the methods currently available it is not possible to solve the LBTE with position-dependent Ga content, we consider uniform Ga impurities over the whole cantilever cross-section. This approximation allows to capture the increased relative influence of the Ga-contaminated layer (of fixed depth $d=0.01\rm{\mu m}$, see main text) when reducing the cantilever width $w$. In particular, we consider an impurity concentration proportional to the ratio of contaminated area to total cross-section area, referred to as penetration ratio $\mathrm{PR}$
%\begin{equation} \label{PR}
	%\mathrm{PR}(w)=\frac{S_{Ga}}{S}\simeq\frac{d}{w}\frac{4}{\sin \theta},
%\end{equation}
%where the approximated equality holds at linear order in $d/w$; details on the geometrical analysis performed to compute PR$(w)$ are reported in the next section.
%Eq.~(\ref{PR}) allows us to describe the width-dependent average Ga concentration $\eta(w)$ as:
%\begin{equation}
	%\eta(w)=\eta_{\mathrm{Ga}}\mathrm{PR}(w),
%\end{equation}
%where $\eta_{\mathrm{Ga}}$ is the concentration of Ga impurities in the contaminated layer.
%We thus determine the conductivity  from the linear-response solution of the LBTE, accounting  for phonon-impurity scattering with impurity strength parameter corresponding to Ga impurities with concentration $\eta(w)$\cite{ratsifaritana1987scattering}.

In summary, we determine the linear-response solution of the LBTE (Eq.~\eqref{lbte}) for diamond nanowires %with width-dependent Ga impurity scattering 
using the \textsf{ShengBTE} package~\cite{li2014shengbte} and use the bulk thermal viscosity computed following the theoretical prescriptions of Ref.~\cite{dragavsevic2023viscous}. Then, we solve the VHE numerically (Eq.~\eqref{VHE}) using a finite-element solver implemented in Mathematica~\cite{mathematica}, imposing a temperature gradient $\Delta T = 10$~K around room temperature on $L=10$~\textmu m long cantilevers. Finally, we obtain the effective thermal conductivity from Eqs~\eqref{heat_flux} and~\eqref{effective_kappa_em}. %The Ga concentration in the contaminated areas $\eta_\mathrm{Ga}$ is set in the \textsf{ShengBTE} simulation so as to fit the experimental trend of $\kappa_{\mathrm{eff}}$ vs width (black squares in Fig. \ref{fig:Fig4} of the main text). 
There is no  fitting parameter in our model.

%Sentence below can be removed
%The resulting theoretical $\kappa_{\mathrm{eff}}$ obtained in this way agrees reasonably with the experimental trend. Hydrodynamics effects (quantified by the difference between the green and red curves in Fig. \ref{fig:Fig4}) do not affect the qualitative behavior of thermal conductivity with varying widths of cantilevers.

\subsection{Estimation of the magnitude of viscous hydrodynamic effects}

\begin{figure}[ht]
\centering
\includegraphics[width=0.65\linewidth]{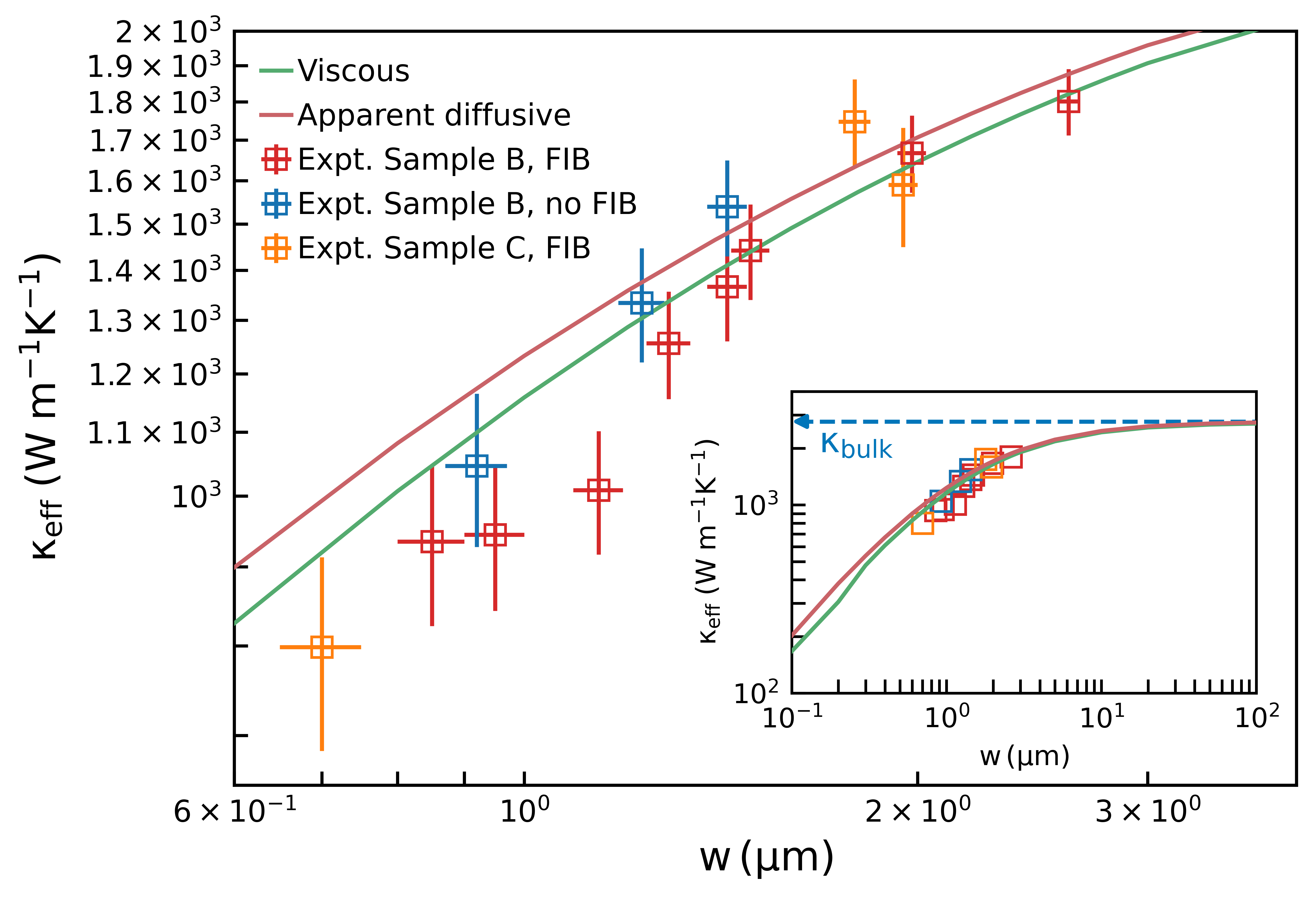}
\caption{
\update{
Effective thermal conductivity of diamond cantilevers as a function of width $w$.
Squares represent experimental data, with error bars reflecting uncertainties in both width and temperature gradient. Solid lines are first-principles predictions based on the VHE (green, Eq.~\eqref{heat_flux}) and in the apparent diffusive case, i.e., Eq.~\eqref{heat_flux_fourier} ~\cite{maassen2015steady} (red), where anomalous non-Fourier size–heat flux relations arise solely from a size-dependent conductivity due to ballistic effects. The dashed line in the inset is the bulk thermal conductivity value.}
}
\label{fig:Fig_SI}
\end{figure}

\update{
Violations from standard diffusive thermal transport can emerge in multiple ways: when microscopic carriers achieve a mean free path comparable to the device size~\cite{allen2018temperature}, or when momentum-conserving phonon-phonon collisions dominate over momentum-relaxing Umklapp collisions~\cite{Simoncelli2020}. Both ballistic and hydrodynamic effects share the common underpinnings of being described by non-local relations between temperature gradient and heat flux~\cite{sendra2022hydrodynamic, gurcan2013transport, dragavsevic2023viscous}. The problem of distinguishing these effects is challenging and not addressed even in pioneering works, which employ phenomenologically parametrized mesoscopic models~\cite{aharon2022direct, palm2024observation} or simplified microscopic models that do not capture hydrodynamic physics~\cite{huang2024graphite}.
Depending on the relative magnitude of hydrodynamic and size effects, we can distinguish different regimes~\cite{maassen2015steady}: (i) “ideal diffusion”, where the thermal conductivity is an intrinsic constant; (ii) “apparent diffusion”, the regime in which anomalous, non-Fourier size-heat flux relations are driven by a conductivity that depends on size due to ballistic effects; (iii) “hydrodynamics”, which originates from the presence of viscous effects dominating over momentum-dissipation effects. }

\update{
To distinguish between these different regimes, we estimate the magnitude of hydrodynamic effects in the cantilevers.
In the limit of negligible hydrodynamic effects, Eqs.~\eqref{VHE} reduce to :
\begin{equation} \label{fourier_law}
	-\kappa\nabla^{2}T(\boldsymbol{R})=0.
\end{equation}
%which, besides having the same functional form, turns out to be intrinsically different from $Q_{i}^{\delta}$, since in this regime $T$ is the only present macroscopic field (see Eq. \eqref{fourier_law}) and therefore is not coupled to $\boldsymbol{u}$ \cite{Simoncelli2020}. 
In the absence of ballistic effects, Eq.~\ref{fourier_law} is exactly Fourier's law of diffusive heat conduction. In the presence of ballistic effects, the thermal conductivity $\kappa$ can no longer be considered an intrinsic material property and depends on size, corresponding to the apparent diffusive regime.
In this regime, the heat flux becomes
\begin{equation} \label{heat_flux_fourier}
	Q_{i}(\boldsymbol{R})=-\kappa\sum_{j}\frac{\partial T(\boldsymbol{R})}{\partial R_{j}},
\end{equation}
The importance of viscous effects can thus be estimated by comparing the results of computing the heat flux with Eq.~\eqref{heat_flux} (full solution) versus Eq.~\eqref{heat_flux_fourier} (apparent diffusive case). 
Fig.~\ref{fig:Fig_SI} shows the effective thermal conductivity computed in both cases, together with experimental data presented in the main text. The two curves show a minor difference over the whole range of width. This means that viscous effects only play a minor role in the observed deviations from pure diffusion. Ballistic effects are thus the main factor in the reduction of $\kappa$ for reducing $w$ in 1-dimensional diamond cantilevers.}

\subsection{Influence of vacancy impurities}

\begin{figure}[ht]
\centering
\includegraphics[width=0.65\linewidth]{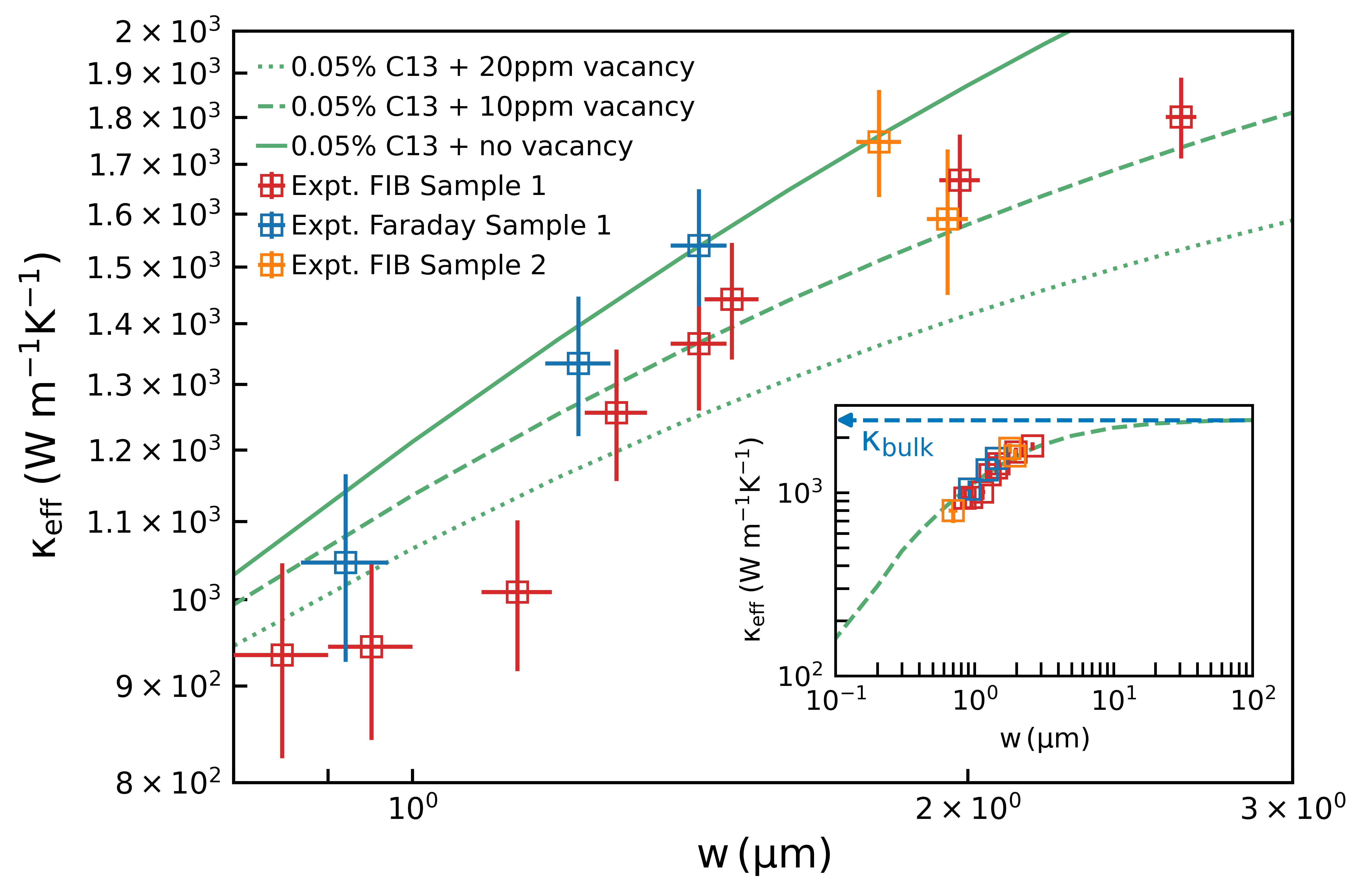}
\caption{
\update{
Effect of vacancy impurities on thermal conductivity of the cantilevers. Solid lines are numerical computation results from the VHE, obtained considering three different impurity content: no vacancies (solid line), 10ppm vacancies (dashed line) and 20 ppm vacancy (dotted line). Squares are exprimental values. The best fit is obtained for 10 ppm vacancies, which is a reasonable estimation for vacancy concentration in our sample, considering the 1-2~ppm NV concentration. The value of 10~ppm is used for Fig.~\ref{fig:Fig4} of the main text.}
}
\label{fig:Fig_SI_impur}
\end{figure}